# Optomechanical cooling and self-stabilization of a waveguide coupled to a whispering-gallery-mode resonator

RICCARDO PENNETTA[*1], SHANGRAN XIE[1], RICHARD ZELTNER[1], JONAS HAMMER[1,2] AND PHILIP ST.J. RUSSELL[1,2]

*[1]Max Planck Institute for the Science of Light and [2]Department of Physics, Friedrich-Alexander-Universität, Staudtstraße. 2, 91058 Erlangen, Germany*

**riccardo.pennetta@mpl.mpg.de*

**Abstract:** Laser cooling of mechanical degrees of freedom is one of the most significant achievements in the field of optomechanics. Here we report for the first time efficient passive optomechanical cooling of the motion of a free-standing waveguide coupled to a whispering-gallery-mode (WGM) resonator. The waveguide is an 8 mm long glass-fibre nanospike, which has a fundamental flexural resonance at $\Omega/2\pi$ = 2.5 kHz and a Q-factor of $1.2\times10^5$. Upon launching ~250 μW laser power at an optical frequency close to the WGM resonant frequency, we observed cooling of the nanospike resonance from room temperature down to 1.8 K. Simultaneous cooling of the first higher order mechanical mode is also observed. The strong suppression of the overall Brownian motion of the nanospike, observed as an 11.6 dB reduction in its mean square displacement, indicates strong optomechanical stabilization of linear coupling between the nanospike and the cavity mode. The combined action of photothermal effects and optical forces between the nanospike and the WGM resonator is identified as the dominant cooling mechanism. The results are of direct relevance in the many applications of WGM resonators, including atom physics, optomechanics and sensing.

## 1. Introduction

Coupling a harmonic oscillator to an optical cavity provides an elegant and powerful means of tailoring its mechanical response [1]. Of particular interest is the regime of "optomechanical cooling", which exploits this coupling to transfer energy from the mechanical motion to the light field, in the process cooling the center-of-mass motion of the mechanical oscillator. In most experimental configurations, this optomechanical coupling is dispersive in nature, i.e., the motion of the harmonic oscillator alters the cavity resonant frequency. Under these conditions optomechanical cooling can be very efficient in the sideband-resolved regime, when the mechanical frequency $\Omega$ is much higher than the linewidth of the optical resonance $\gamma$, i.e., $\gamma \ll \Omega$. For the best optical cavities, $\gamma$ normally lies in the range of a few hundreds of kHz [2]. Increasing the mass or dimensions of a mechanical system implies, however, a decrease in the resonant frequency and consequently inefficient cooling rates [1]. Nevertheless, numerous optomechanical systems operate at low resonant frequencies, for example the mirrors used in LIGO [3], ultracold atomic gases [4] and suspended micro-mirrors [5]. In this regime several alternative cooling schemes have been proposed and demonstrated over the last decade [6-9].

To date optomechanical cooling remains one of the very few ways to manipulate the noise spectrum of mechanical oscillators, whose large dimensions make it difficult to use cryostats or whose thermal coupling with the surrounding environment is weak (i.e. optically levitated particles [10] or long suspended waveguides [11, 12]). In the second category, tapered optical nanofibers play a pivotal role, offering an effective means of interfacing with fiber-based networks and coupling light into photonic devices such as whispering-gallery-mode (WGM) resonators [13], photonic crystal cavities and photonic circuits [14]. Furthermore, recent investigations suggest that understanding and control of the mechanical resonances of tapered nanofibers could be highly beneficial for experiments in atomic physics [15, 16], sensing [17, 18] and optomechanics [19]. Due to the low thermal conductivity of glass, placing a tapered fiber into a cryostat in high vacuum does not suffices to cool the temperature of the fiber waist and more involved solutions are required [20].

In this article, we report strong passive optomechanical cooling of a tapered glass-fibre "nanospike" coupled to a WGM bottle-resonator. Despite several notable reports describing optomechanical interactions between a suspended waveguide coupled to a WGM resonator [19, 21, 22], so far passive cooling has evaded experimental observation, mostly due to the difficulty of obtaining high-Q mechanical modes for the waveguide without compromising its optical properties. We report that appropriately tapered glass-fiber nanospikes offer an elegant solution, providing both adiabatic guidance of light and flexural resonances with quality factors $Q > 10^5$ [23, 24]. This represents an increase of two to three orders of magnitude with the values reported for traditional optical nanofibers [19]. When the nanospike is placed close to a bottle-resonator and the pump laser is blue-detuned from the optical cavity resonance, clear optomechanical cooling is observed. The result is strong suppression of Brownian nanospike motion, indicating self-stabilized coupling to the WGM resonator.

## 2. Experimental setup

The experimental setup is sketched in Fig. 1a. The nanospike was fabricated by scanning an oxybutane flame along a length of single mode step-index fiber (SM980) while gently pulling it. The profile of the nanospike was engineered to yield single-mode adiabatic guidance of light at 1150 nm (pump laser) and 1064 nm (probe 1), while preserving high mechanical stiffness and low mechanical loss. The procedure to reproducibly fabricate nanospikes with high-Q flexural resonance has been previously reported in details [23-25]. The experiment was conducted in vacuum ($10^{-5}$ mbar) to eliminate viscous damping by air. The nanospike was mounted on a stainless-steel holder using Kapton tape. Carefully cleaning of the optical fibre before tapering ensures that the subwavelength waist can withstand several tens of mW of optical power in high vacuum without any damage. Five stepper motors permitted fine-tuning of both the relative position and the orientation between the nanospike and the WGM bottle-resonator. An optical micrograph of nanospike+bottle-resonator is shown in Fig. 1b. The resonator [26, 27] was fabricated in a two-step process. First a piece of single mode fibre (SMF) was thermally tapered to a diameter of ~20 µm, and then the taper was placed in an arc-splicer, where an electric discharge locally heated the taper waist, while the two ends of the taper were pushed towards each other. Surface tension caused the formation of the prolate shape shown in Fig. 1b. Tuning the arc power and its duration allowed precise control of the resonator diameter and profile. In the vicinity of a cavity resonance the pump light transmission is low, making it difficult to image the motion of the nanospike tip. To side-step this problem, light from a second weak and non-resonant probe laser (probe 1 in Fig. 1a) was launched into the fiber so as to allow the nanospike motion to be monitored using a quadrant photodiode (QPD). This permitted the tip motion to be reconstructed in two dimensions with nm-scale spatial resolution. To calibrate the displacement measured by QPD, the nanospike was moved using the motors inside the vacuum chamber with its displacement precisely measured using the top CCD (see Fig. 1).

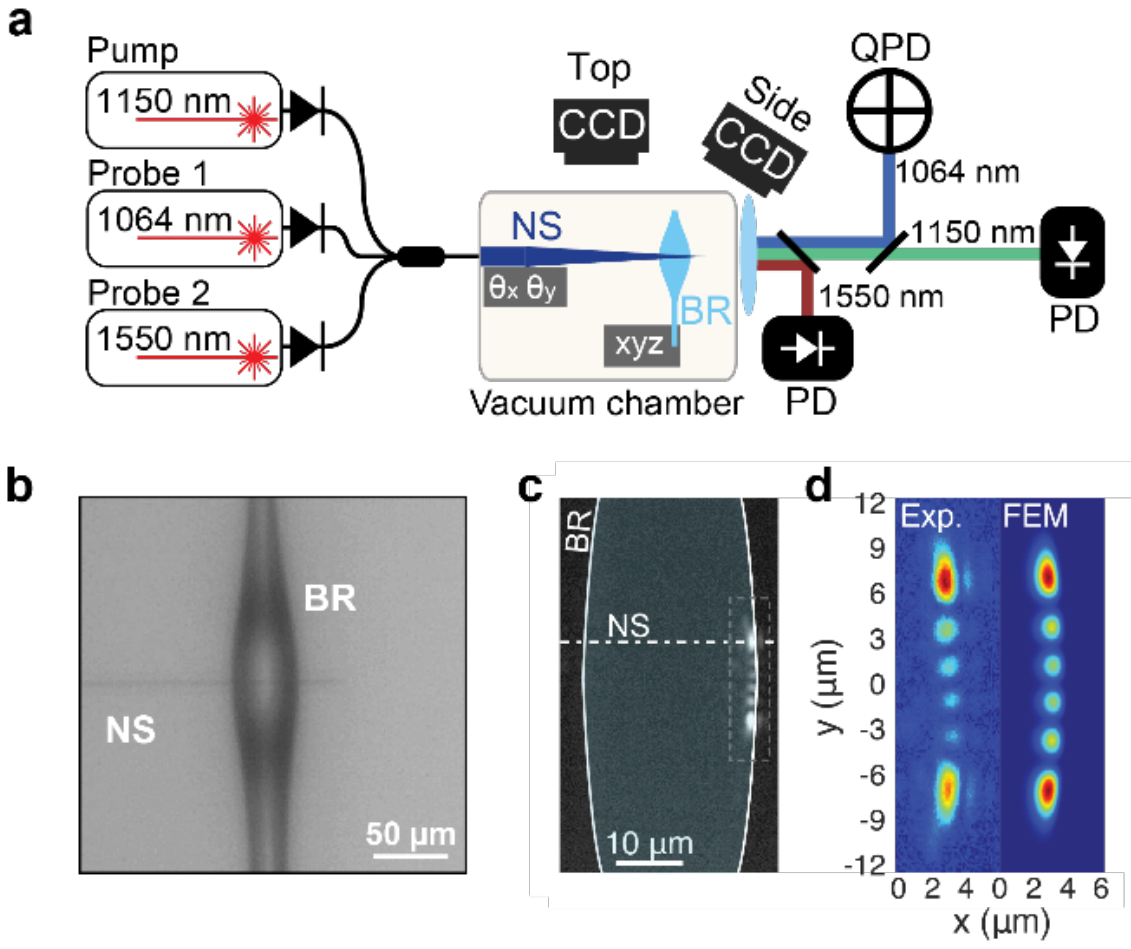

Fig. 1. Nanospike coupled to a WGM bottle resonator. (a) Sketch of the experimental set-up. NS, nanospike; BR, Bottle resonator, QPD, quadrant photodiode, PD, photodiode. (b) Optical micrograph of the nanospike coupled to a WGM resonator from the side c) Micrograph of a bottle resonator when the laser light is locked into resonance, light being launched from the right. The weak signal radiated by the bottle resonator on resonance could be used to image the near-field of the optical mode using a microscope objective and a sensitive NIR camera. (d) Zoom-in of one of the measured optical mode profiles compared to the result of finite element simulations.

## 3. Optomechanical cooling of nanospike motion

The first set of experiments was performed using a nanospike with tip diameter ~500 nm and a fundamental flexural resonance at 2.5 kHz (Q-factor of $1.2\times10^5$). The bottle-resonator had a diameter of 46 µm and an intrinsic optical Q-factor of $8.2\times10^7$. Fig. 1c shows the near-field intensity distribution of the optical mode, measured when the pump beam was locked on resonance and captured by a sensitive CCD camera. The measurement matches well with the results of finite element simulations (Fig. 1d).

Strong optomechanical cooling was observed when the nanospike was placed in the over-coupled regime with the pump wavelength set very close to, but blue-detuned from, the cavity resonance. The laser detuning was stabilized using the thermo-optical nonlinearity of glass, i.e., by thermal self-locking [28]. The resulting mechanical spectra in the vicinity of the fundamental nanospike resonance are depicted in Fig. 2a for increasing values of pump power. A significant drop in the amplitude of the mechanical resonance, accompanied by linewidth broadening, is apparent at higher pump powers. The effective temperatures ($T_{\mathrm{eff}}$) of the nanospike "degree of freedom" were estimated by integrating the area underneath the power spectra [1]. As shown in Fig. 2b, an increase of several orders of magnitude in the mechanical linewidth could be measured at only 250 µW pump power, with a minimum $T_{\mathrm{eff}}$ value of 1.8 K. Note that these results refer to nanospike motion orthogonal to the WGM resonator surface; weaker optomechanical coupling is expected in the direction parallel to the WGM resonator surface. Nonetheless this degree of freedom could still be cooled, as shown in the insets of Fig. 2a and Fig. 2b, with a minimum achievable effective temperature of 68 K. The saturation of $T_{\mathrm{eff}}$ at higher pump power observed in Fig. 2b may be related with the residual optical absorption in the bottle-resonator, however a more precise understanding of its origin requires further investigations.

Since the nanospike mechanical frequencies are much smaller than the cavity linewidth, the cooling should not differ substantially for higher order mechanical modes [1]. Fig. 2c shows the measured power spectra of the first higher order flexural mode, using the pump power as a parameter (same data set as in Fig. 2a). At zero pump power, this mode had a resonance frequency of 6.45 kHz and a Q-factor of 1840. The clear trend observed when increasing the power of the pump laser confirms simultaneous cooling of the higher order mode. Because of the lower mechanical Q-factor, the minimum achievable $T_{eff}$ was 118 K (inset of Fig. 2c). It is worth mentioning that multimode cooling is not easily achievable in the sideband-resolved regime because laser detuning must be precisely matched to the mechanical resonant frequency.

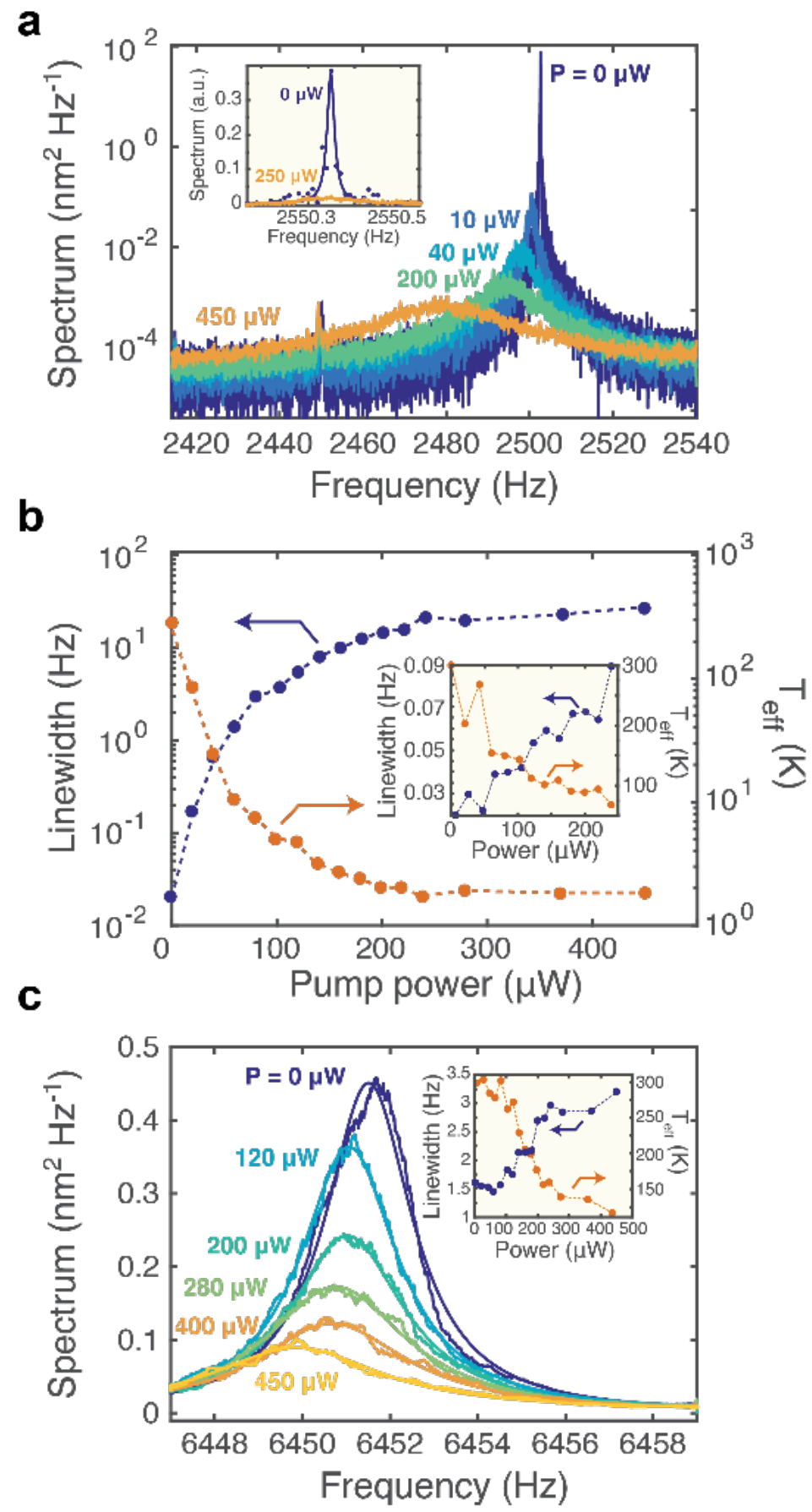

Fig. 2. Optomechanical cooling of nanospike motion. (a) Measured mechanical power spectrum in the vicinity of the fundamental (flexural) nanospike mode for 5 different pump powers. Inset: measured power spectrum for vibrations parallel to the surface (see text) for 0 and 250 μW pump power. The solid lines are Lorentzian fits. (b) Mechanical linewidth (left axis) and inferred effective temperature Teff (right axis) as a function of pump power. The dashed lines are guides for the eye. Inset: same measurement as in (b) but for vibrations parallel to the surface. (c) Measured mechanical power spectra in the vicinity of the first high order (flexural) nanospike mode for increasing values of pump power. The solid lines are Lorentzian fits. Inset: linewidth (left axis) and effective temperature Teff (right axis) of the same mechanical mode as a function of the pump power.

## 4. Self-stabilized coupling to WGM resonator

Laser cooling of the first few flexural mechanical modes of the nanospike resulted in strong stabilization of its coupling to the WGM resonator. At room temperature in the absence of stabilization, Brownian motion of the nanospike causes fluctuations as high as tens of nm in its position. This causes random fluctuations in the frequency and linewidth of the cavity resonance as well as the optical transmission through the nanospike. Fig. 3a plots the displacement of the nanospike (after calibrating the response of the QPD) recorded over 100 ms for pump powers of 0 μW, 10 μW and 250 μW. The panel on the right-hand side compares histograms for data collected over 100 s. The reduction in the thermal noise can be clearly observed. At low power, the Brownian motion of the nanospike has a mean-square displacement (MSD) of 530 nm$^2$, in agreement with estimates from the equipartition theorem. The effective mass (estimated by solving the Euler–Bernoulli equation) referring to the displacement of the tip of the nanospike is $m_{\mathrm{eff,FM}} \approx 50$ pg for the fundamental mechanical mode and $m_{\mathrm{eff,HOM}} \approx 70$ pg for the first higher order mode. When the pump power was increased to 250 μW, the value of the MSD drops significantly to 37 nm$^2$ (Fig. 3b) – a suppression factor of 11.6 dB. At this level, the MSD is dominated by the higher order mechanical modes.

The effect of stabilization was further explored by introducing a second frequency-tunable 1550 nm probe laser, (probe 2 in Fig. 1a), and scanning its wavelength across the cavity resonances with and without laser cooling. Fig. 3d and 3e compare 50 consecutive measurements of the resonance for a $T_{\mathrm{eff}}$ of 300 K and 6.7 K, revealing a clear overall increase in the system stability. In particular, nanospike cooling significantly reduced the measured standard deviation of the minimum transmission from $\sigma_{300\mathrm{K}} = 0.0117$ to $\sigma_{6.7\mathrm{K}} = 0.0025$ (see Fig. 3f and 3g).

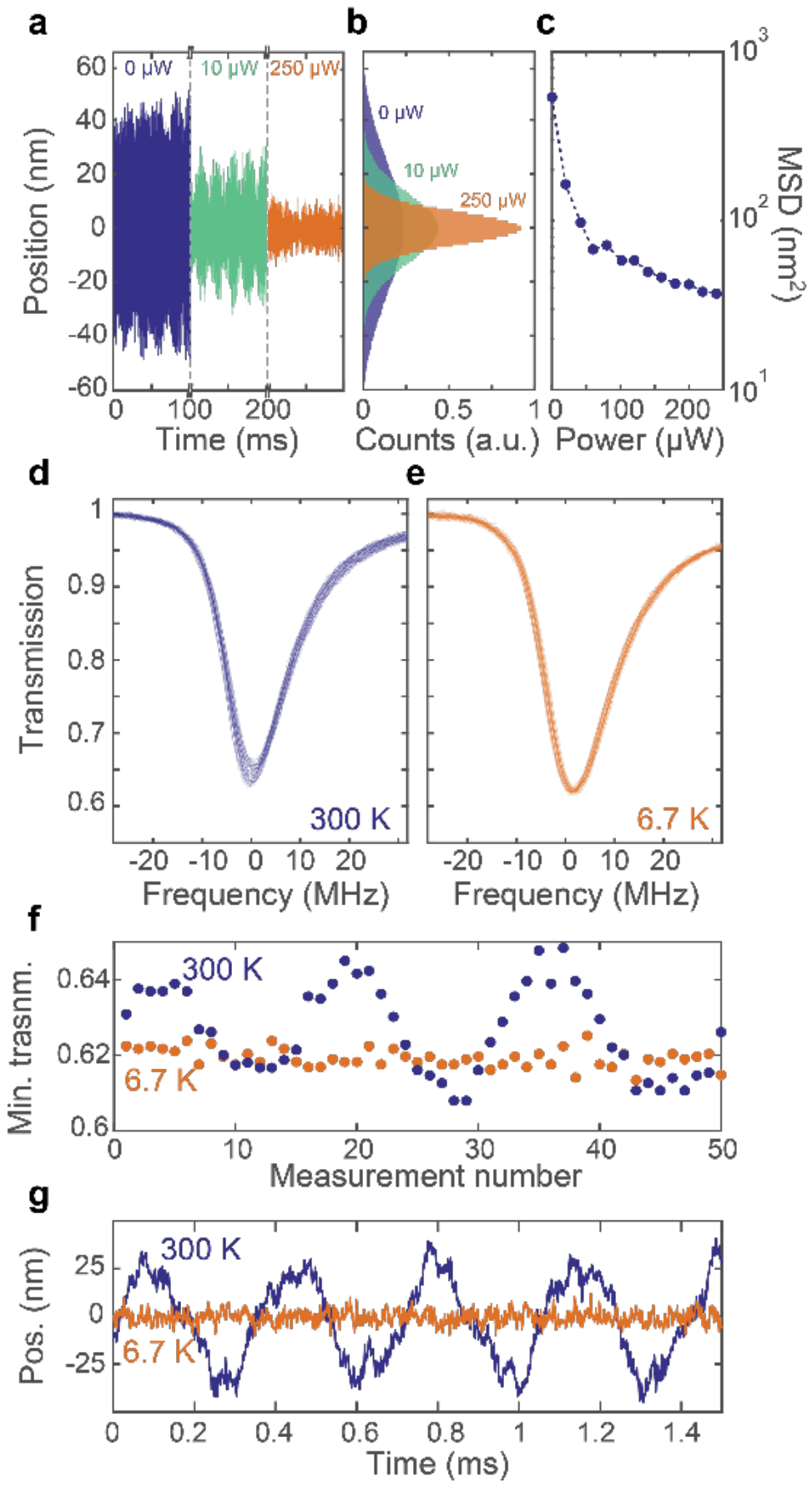

Fig. 3 (a) Temporal motion of the nanospike for different launching pump powers (the sampling rate is 20 kHz). (b) Histogram plots of the nanospike displacements. (c) Mean-squared displacements of the nanospike as a function of pump power. (d, e) 50 consecutive measurements of a cavity resonance observed with the second probe laser (1550 nm) when $T_{\mathrm{eff}}$ equals 300 K (c) and 6.7 K (d). (f) Minimum transmission recorded in (d) and (e) as a function of the measurement number; the blue dots correspond to $T_{\mathrm{eff}}$ = 300 K, and the orange dots to $T_{\mathrm{eff}}$ = 6.7 K. The fluctuations in the experimental data are artifacts of the short total acquisition time (1 s), which was much less than the lifetime of the fundamental mechanical mode (~30 s). (g) Nanospike deflection collected over 1.5 ms for $T_{\mathrm{eff}}$ = 300 K (blue line) and $T_{\mathrm{eff}}$ = 6.7 K (orange line).

## 5. Cooling mechanism

Nanospike motion modulates both the resonant frequency and the decay rate of the optical cavity, introducing simultaneously dispersive and dissipative optomechanical coupling [6]. In addition, since the experiment was performed in vacuum, photothermal effects due to residual absorption of the pump power can also contribute to the observed cooling effect [9].

Measuring the frequency and linewidth of the optical mode as a function of the distance between the nanospike and the bottle-resonator allows the dispersive and dissipative optomechanical coupling parameters to be estimated (see Appendix B for the data). Comparing the results with the theory of generalized optomechanical coupling [6], however, we found that the measured cooling efficiency exceeded the predicted values by about 2 orders of magnitude, suggesting that photothermal interactions were strongly affecting the cooling process. In particular, since the bottle resonator takes a finite time (time constant $\tau$ = 280 μs, see Appendix D) to reach thermal equilibrium, thermal nonlinearities delay the build-up of optical energy in the cavity, producing non-conservative optical forces and strongly perturbing the optomechanical state.

To further clarify the cooling mechanism, we performed measurements using a bottle-resonator with a larger diameter of 350 μm, the aim being to suppress photothermal effects. The increase in volume results in a greater heat capacity and, since the fabrication procedure does not require pre-tapering, a reduction in water diffusion into the glass and consequently less residual laser absorption. In addition, a greater thermal response time of $\tau$ = 8.3 s (about 40 times longer than in the smaller bottle resonator) was measured, which should further suppress photothermal coupling. An increase in the intrinsic optical Q-factor to a value of $2.6\times10^{8}$ was also observed.

The second set of experiments was performed using a nanospike with resonant frequency 1.9 kHz (Q-factor of $1.4\times10^{5}$) and a tip diameter of ~700 nm ($m_{\mathrm{eff}} \approx$ 780 pg). Under these circumstances, as shown in Fig. 4a, the measured optical spring effect for a launched power of 70 μW (blues points) agrees very well with the predictions of the model of generalized optomechanical interactions (solid-blue line, see Appendix C for the parameter list). The model also correctly estimates the pump power (~300 μW) required to access the regime of mechanical self-oscillation for the fundamental mechanical mode orthogonal to the WGM surface (inset of Fig. 4a, with cavity detuning of 1.6 MHz).

The good agreement between experiment and theory suggests that photothermal effects have little relevance in this parameter range and that the generalized optomechanical model correctly describes the system dynamics. This also suggests that the results in Figs. 2 and 3 are dominated by photothermal effects.

Using the same parameters for the measurement in Fig. 4a (launched pump with power 1 mW, blue-detuned by 0.2 MHz from the cavity resonance), dissipative cooling of the fundamental mechanical mode to an effective temperature of 46 K is predicted. Unfortunately, the very large mechanical frequency shift due to the optical spring effect at this power level (~65 Hz, orders of magnitude larger than the mechanical linewidth), together with a significant drop in the efficiency of the thermal self-locking mechanism in this bottle-resonator, prevented collection of clean mechanical spectra and verification of the theoretical predictions.

## 6. Mode coupling and anti-crossing

Vibration of the nanospike in one of its mechanical eigenmodes modulates the optical field in the bottle-resonator and drives the other mechanical modes via the optical force. Since all the mechanical modes of the nanospike interact with the same cavity mode, they become optomechanically coupled [29]. Though this coupling can be neglected in systems where the eigenmodes have widely different resonant frequencies, this is not true in our case because the orthogonal nanospike modes are nearly degenerate in frequency. Using the optical spring effect, we were able to experimentally characterize the strength of the optomechanical coupling. The results are presented in Fig. 4b, where the measured mechanical frequencies for the two fundamental flexural modes orthogonal and parallel to the bottle-resonator surface are shown as a function of the cavity detuning at a power level of 40 μW. A clear anti-crossing can be observed at 4 MHz detuning, with a measured coupling rate of 1.8 Hz – two orders of magnitude greater than the mechanical decay rate. The solid lines in Fig. 4b are fits of the data to coupled mode theory (see Appendix E), showing good agreement with the measurements.

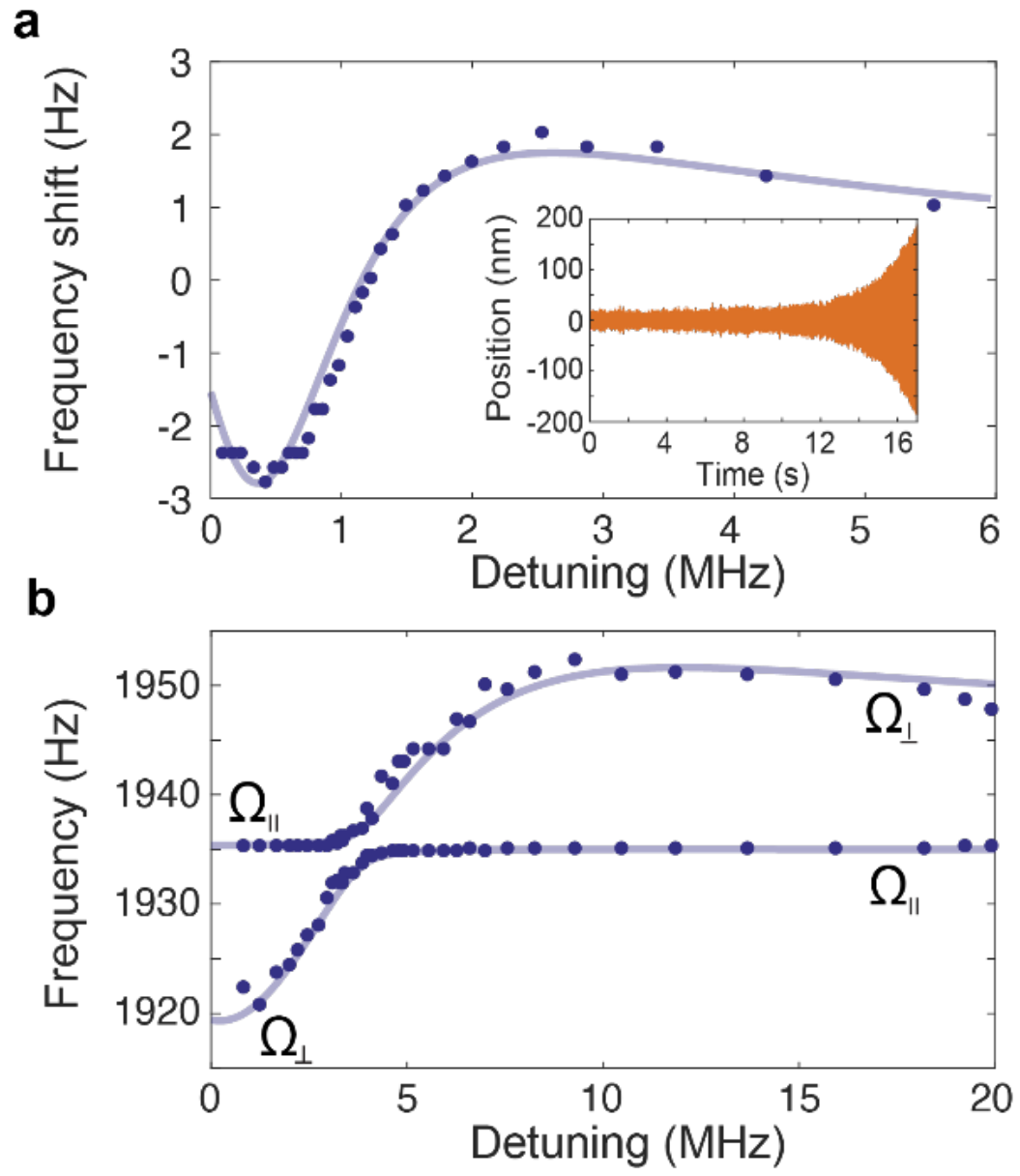

Fig. 4 (a) Measured frequency shift of the fundamental flexural mode of the nanospike coupled to a WGM resonator with a diameter of 350 μm, plotted against laser detuning for a launched power of 70 μW (blue dots). The solid line is a fit to the model for generalized optomechanical coupling. Inset: Nanospike deflection as a function of time when the pump power was raised just above the threshold for self-oscillation for a laser detuning of 1.6 MHz. (b) Measured mechanical frequency as a function of laser detuning for mechanical oscillation of the nanospike parallel ( $\Omega_p$ ) and orthogonal ( $\Omega_{\perp}$ ) to the WGM surface (see Appendix E for detailed information).

## 7. Conclusions

In summary, glass-fibre nanospikes permit observation for the first time of passive cooling of a free-standing optical waveguide evanescently coupled to a WGM resonator. Accurate knowledge of the position and velocity of the mechanical resonator is not required, in sharp contrast to active cooling schemes [19], which suffer from noise at very low motion temperature under which condition the amplitude of the mechanical oscillations is greatly reduced. We demonstrate that cooling of the nanospike motion is highly beneficial for stabilizing the coupling to an optical cavity well beyond the limits set by thermodynamics, a result that is highly relevant in numerous applications of WGM resonators [16-18, 30]. Moreover, in this configuration, optical cooling happens with blue-detuned laser light, for which the cavity is intrinsically stable [28]. Finally, this approach is general and relatively simple and may be applied to any type of optomechanical system that has a high enough mechanical Q-factor, allowing efficient laser cooling of low-frequency mechanical oscillators.

## Appendix A: Generalized optomechanical coupling

A theoretical model for generalized optomechanical coupling is described in [6]. The dispersive and dissipative parts of the optomechanical coupling come respectively from the position dependence of the cavity resonance frequency $\omega_C$ and the decay rate $\gamma_{\text{ext}}$ of the cavity resonance through coupling to the input waveguide. For small oscillation around an equilibrium position $x_0$, the optomechanical interactions can be quantified by approximating the parameters $\theta_1$ (dispersive coupling) and $\gamma_1$ (dissipative coupling), as linear functions of nanospike deflection $x$:

$$\theta(x) = (\omega_L - \omega_C) + \theta_1 x \tag{A1}$$

$$\gamma_{\text{ext}}(x) = \gamma_{\text{ext}}(x_0) + \gamma_1 x \tag{A2}$$

where $\omega_L$ the laser frequency. It is possible to show that under these conditions the optical cavity acts back on the mechanical resonator, modifying its linewidth $\Gamma$ and resonant frequency $\Omega$. Explicit analytical expressions for the corresponding correction factors $\delta\Gamma_{\text{opt}}$ and $\delta\Omega_{\text{opt}}$ can be found in [6]. The expected effective temperature $T_{\text{eff}}$ of the system can be estimated using the following relation [31]:

$$T_{\text{eff}} = T\frac{\Gamma}{\Gamma + \delta\Gamma_{opt}} \tag{A3}$$

where $T$ is the ambient temperature.

We note that in the most common optomechanical systems (e.g. Fabri-Perot microcavities, optomechanical crystals, mechanical WGM resonators) the optomechanical coupling is purely dispersive. Dissipative coupling was first theoretically investigated in Ref. [6] and since then experimentally demonstrated in a relatively limited number of systems including nanomechanical beam waveguides [21], Michelson-Sagnac interferometers [7] and split-beam nanocavities [32].

## Appendix B: Optomechanical coupling parameters

The values of the parameters $\theta_1$ and $\gamma_1$ can be estimated by measuring the cavity resonance frequency and linewidth as a function of the distance between the nanospike and the bottle-resonator. To illustrate this, Fig. 5a shows the results for a bottle-resonator with diameter 259 μm. The measurement was performed at atmospheric pressure and a laser power of 0.5 μW. The solid lines are exponential fits to the experimental data. The coupling parameters (Fig. 5b) were then calculated as a derivative of the fits.

Repeating the measurement for different optical modes, we found consistent values for $\gamma_1$, although both the sign and the magnitude of $\theta_1$ were found to change significantly. The situation did not change when the measurement was repeated for other bottle-resonators. As an example, in Fig. 6b we show the measured value $\theta_1/\gamma_1$ for bottle-resonators with significantly different diameters and for several optical modes.

As a side-note, as shown in Fig. 6c the optical Q-factor was observed to increase with the diameter of the bottle resonator.

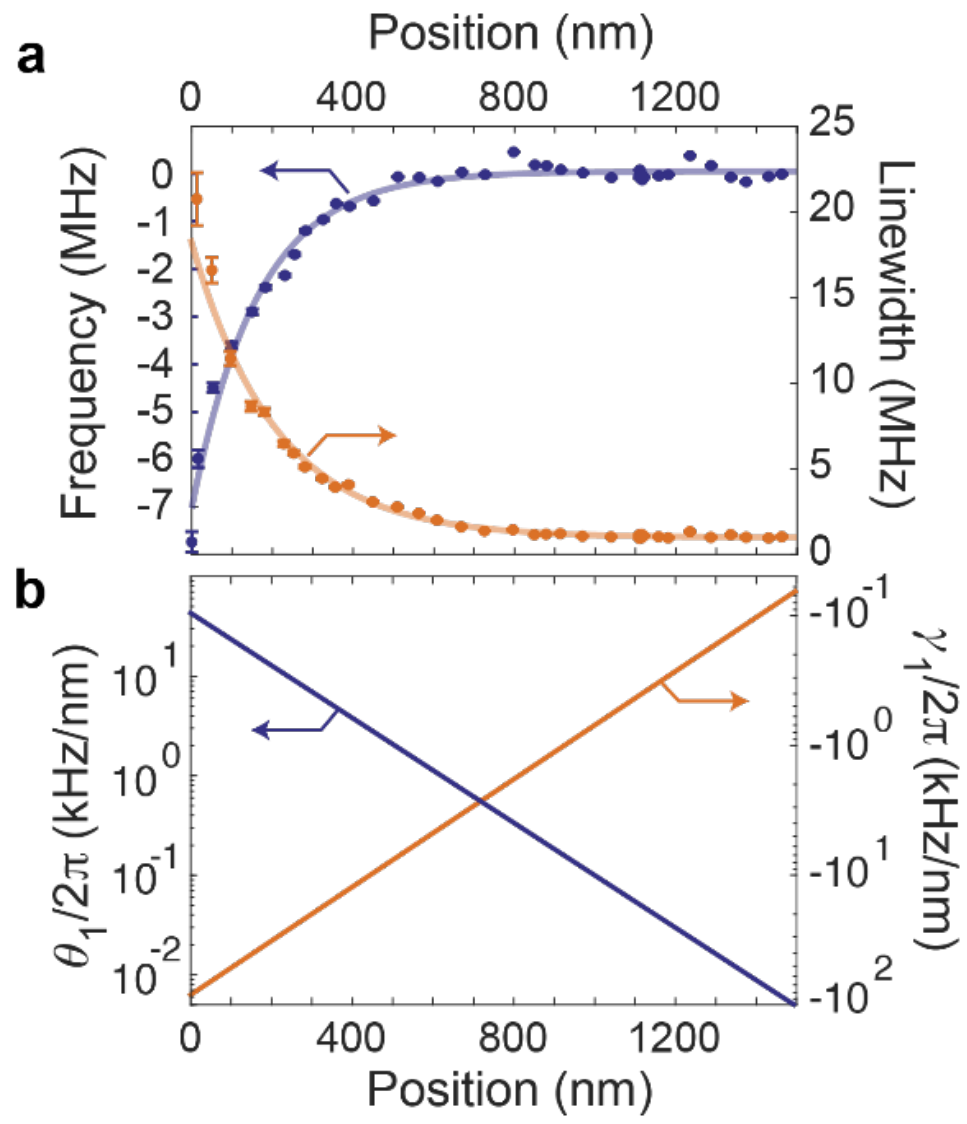

Fig. 5 (a) Measured frequency shift (left axis) and linewidth (right axis) for the excited WGM plotted against nanospike position. The solid lines are fits to the data using exponential functions. (b) Coupling parameters $\theta_1$ (dispersive coupling, right axis) and $\gamma_1$ (dissipative coupling, right axis) plotted as a function of nanospike position.

## Appendix C: System parameters

The data presented in Fig. 2 and 3 have been collected using a nanospike with a resonant frequency $\Omega/2\pi = 2.5$ kHz and a Q-factor of $1.2\times10^5$. The effective mass is $m_{\mathrm{eff,FM}} \approx 50$ pg for the fundamental mode and $m_{\mathrm{eff,HOM}} \approx 70$ pg for the first higher order mode The bottle resonator had a diameter of 46 μm and an intrinsic optical Q-factor of $8.2\times10^7$.

The results in Fig. 4 have been recorded using a nanospike with a resonant frequency $\Omega/2\pi = 1.9$ kHz and a Q-factor of $1.4\times10^5$. The bottle resonator used had a diameter of 350 μm. In particular, the parameters used to formulate the theoretical prediction of Fig. 4a are as follows:

Laser wavelength $\lambda = 1150$ nm
Optical power $P = 70$ μW
Internal WGM loss $\kappa_{\mathrm{int}} = 2\pi\times1.0$ MHz
External WGM loss $\kappa_{\mathrm{ext}} = 2\pi\times1.3$ MHz
Mechanical frequency $\Omega = 2\pi\times1928$ Hz
Effective mass $m_{\mathrm{eff}} = 780$ pg

The coupling parameters were obtained by fitting the model of Ref. [6] to the experimental data. The resulting values are: $\gamma_1 = 2\pi\times7.1$ kHz/nm, $\theta_1 = 2\pi\times0.71$ kHz/nm. Note that these values lie well within the parameter range expected for the system, as shown in Fig. 6. Unfortunately, it was not possible to measure the coupling parameters inside the vacuum chamber because of the lack of calibrated vacuum-compatible motors. However, we found good agreement between the fitted values and the measurement in Fig. 5.

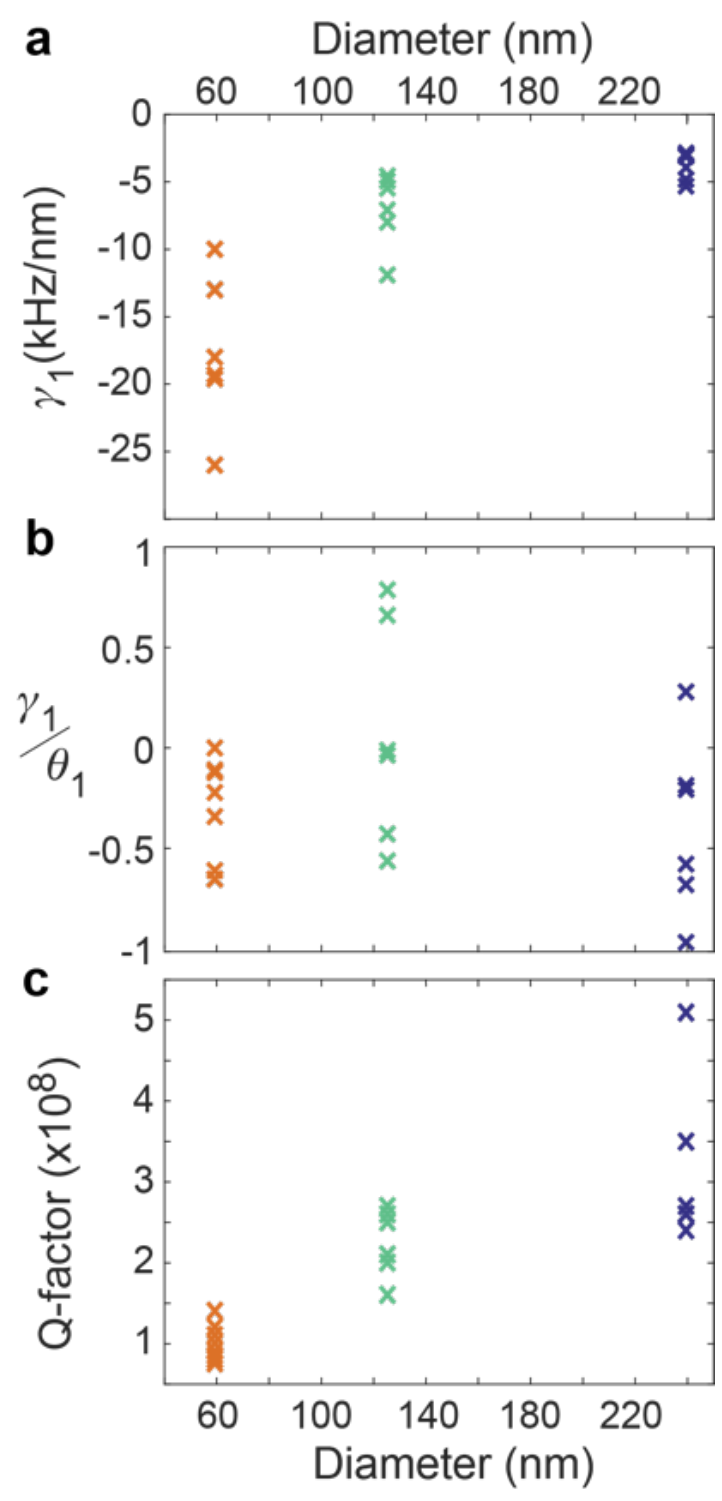

Fig. 6 (a) Dissipative coupling parameter $\gamma_1$ measured at critical coupling, (b) corresponding ratio between the dissipative and dispersive coupling parameters (i.e. $\gamma_1/\theta_1$) and (c) intrinsic Q-factor for bottle-resonators with different diameters. Each cross corresponds to a different optical mode.

## Appendix D: Thermal response time

Cooling of mechanical motion via photothermal coupling was first demonstrated in [9]. A key feature in this scenario is that the typical thermal response time $\tau$ of the system should not differ too much from the inverse of the mechanical resonance frequency. We measure the thermal response time of the WGM resonators under high vacuum using a two-color scheme. One laser (probe 1 in Fig. 1a) was locked to a WGM, resulting in slight heating of the resonator, while a second low power laser (labelled pump in Fig. 1a) was scanned across another resonance to measure its spectral position. After quickly (< 10 ms) switching off the first laser, the drift in the cavity resonance monitored by the second laser was recorded as a function of time. The results obtained from resonators with diameters of 350 μm and 40 μm are shown in Fig. 7. By fitting the data to exponential functions [9] we obtained thermal time constants: $\tau_{40\mu m} = 0.28$ s and $\tau_{350\mu m} = 8.3$ s.

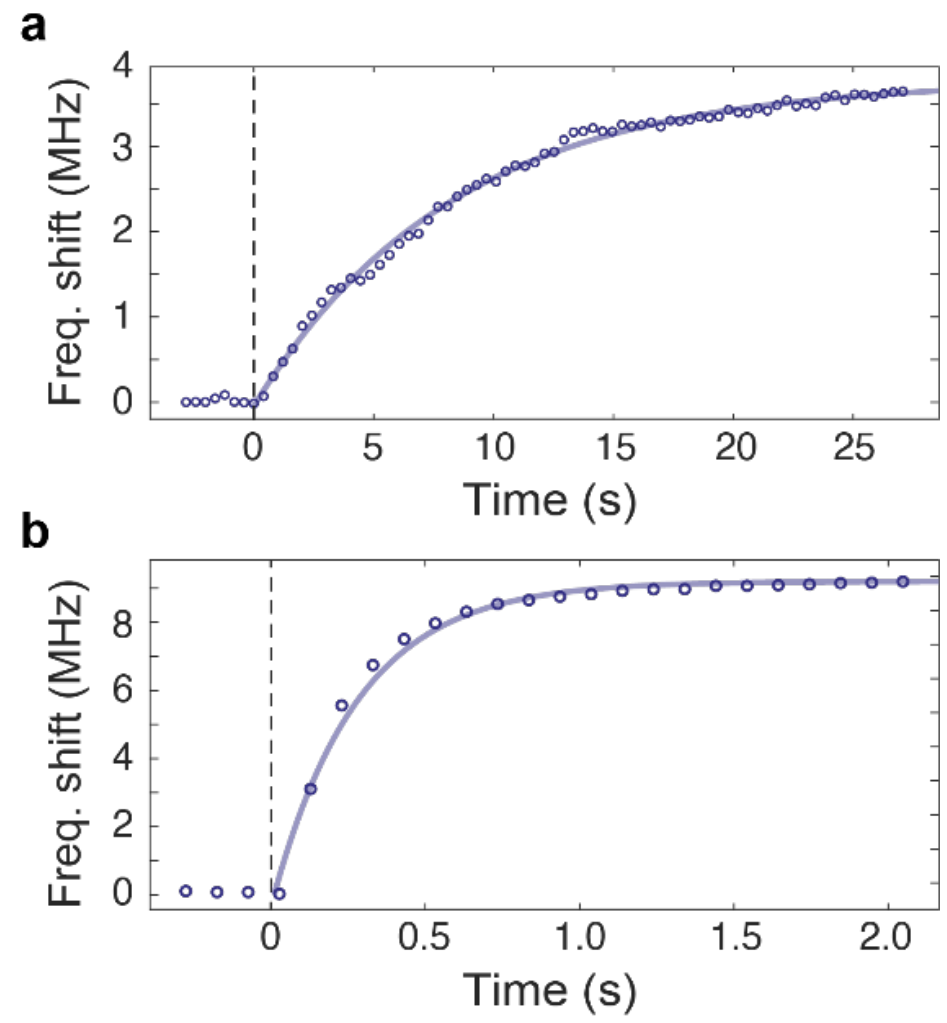

Fig. 7 Measured resonant frequency shift over time for bottle-resonators with diameters (a) 350 μm and (b) 40 μm. Fitting the data to exponential functions (solid lines) yields thermal time constants (a) $\tau_{350\mu m}$ = 8.3 s and (b) $\tau_{40\mu m}$ = 0.28 s.

## Appendix E: Optically coupled mechanical modes

In the Appendices A and B a simple unidirectional model was used to describe the motion of the nanospike. Some of the features observed in the experiment require, however, a more general treatment. The optical force acting on the nanospike when placed in the proximity of the bottle-resonator can be estimated, to a first approximation, as proportional to the gradient of the local intensity I:

$$\vec{F} \propto \vec{\nabla} I \tag{A4}$$

For small displacements around an equilibrium position and considering a Cartesian frame of reference whose *x*-axis is orthogonal to the surface of the bottle-resonator, the components of the optical force can be approximated as linear functions of position:

$$\begin{aligned} F_x(x,y) &\approx -k_{xx}x - k_{xy}y \\ F_y(x,y) &\approx -k_{yx}x - k_{yy}y \end{aligned} \tag{A5}$$

where $k_{ij} = \partial F_i/\partial j$ are the optical stiffness parameters and $k_{xy} = k_{yx}$ from Eq. A4.

Under these conditions and neglecting mechanical damping (very weak in our system), the two-dimensional motion of the nanospike can be described using the following set of equations:

$$\begin{aligned} \ddot{x} + \Omega_x^2 x &= -\frac{k_{xx}}{m_{eff}}x - \frac{k_{xy}}{m_{eff}}y \\ \ddot{y} + \Omega_y^2 y &= -\frac{k_{xy}}{m_{eff}}x - \frac{k_{yy}}{m_{eff}}y \end{aligned} \tag{A6}$$

where $\Omega_x$ and $\Omega_y$ are the intrinsic resonant frequencies for the mechanical modes along these two orthogonal directions and are nearly degenerate in our experiment ($\Omega_x/\Omega_y \approx 1$). We solve Eqs. A6 introducing slowly varying envelopes for x(t) and y(t):

$$\begin{aligned} x(t) &= \chi(t)\, e^{i\bar{\Omega}t} \\ y(t) &= \eta(t)\, e^{i\bar{\Omega}t} \end{aligned} \tag{A7}$$

where $\bar{\Omega} = (\Omega_x + \Omega_y)/2$. Inserting A7 into A6, neglecting second order derivates for *χ(t)* and *η(t)*, and cosidering that $(\Omega_x + \bar{\Omega}) \approx (\Omega_y + \bar{\Omega}) \approx 2\bar{\Omega}$, the equations of motion can be rewritten as:

$$\begin{aligned} \dot{\chi} - i\frac{\delta\Omega}{2}\chi &= -i(\kappa_{xx}\chi + \kappa_{xy}\eta) \\ \dot{\eta} + i\frac{\delta\Omega}{2}\eta &= -i(\kappa_{xy}\chi + \kappa_{yy}\eta) \end{aligned} \tag{A8}$$

Where $\delta\Omega = \Omega_x - \Omega_y$ and $\kappa_{ij} = k_{ij}/(2m_{eff}\bar{\Omega})$. Solving for the natural frequencies of the system and considering Eqs. A7 we obtain the resonant frequencies of the coupled mechanical modes:

$$\Omega_{\pm} - \bar{\Omega} = \frac{\kappa_{xx} + \kappa_{yy}}{2} \pm \sqrt{\left(\frac{\delta\Omega}{2} - \frac{\kappa_{xx} + \kappa_{yy}}{2}\right)^2 + \kappa_{xy}^{\,2}} \tag{A9}$$

In our system, since the intensity gradient is steeper along the *x*-axis: $\kappa_{yy} \ll \kappa_{xy} \ll \kappa_{xx}$ and it is therefore possible to neglect the contribution of $\kappa_{yy}$. Also $\kappa_{xx}$ correspond to the optically induced frequency shift if a single degree of freedom is considered and it can be estimated as discussed in Appendix A (i.e. $\kappa_{xx} = \delta\Omega_{opt}$). The two solutions of Eq. A9 are represented as solid lines in Fig. 4b, using the parameters:

| | |
|---|---|
| Coupling coefficient | $\kappa_{xy}$ =1.8 Hz |
| Laser wavelength | $\lambda$ = 1150 nm |
| Optical power | P = 40 μW |
| Internal WGM loss | $\kappa_{int} = 2\pi\times 1.5$ MHz |
| External WGM loss | $\kappa_{ext} = 2\pi\times 12$ MHz |
| Mechanical frequencies | $\Omega_x$ = 1945 Hz |
| | $\Omega_y$ = 1935 Hz |
| Effective mass | $m_{eff}$ = 780 pg |
| Dissipative coupling | $\gamma_1 = 2\pi\times 88.0$ kHz/nm |
| Dispersive coupling | $\theta_1 = 2\pi\times 35.2$ kHz/nm |

## Funding

Max Planck Society

## Acknowledgments

We thank G. Epple, F. Marquardt, F. Sedlmeir, J. Harris, A. Kashkanova and A. Shkarin for useful discussions.

## Disclosures

The authors declare no conflicts of interest.